\documentclass[12pt]{article}
\usepackage{graphicx}

\def\Re{{\cal R \mskip-4mu \lower.1ex \hbox{\it e}\,}}
\def\Im{{\cal I \mskip-5mu \lower.1ex \hbox{\it m}\,}}

\def\eg{{\it e.g.}}

\def\etal{{\it et al.}}

\def\sub#1{_{\lower.25ex\hbox{$\scriptstyle#1$}}}
\def\tev{\,{\ifmmode\mathrm {TeV}\else TeV\fi}}
\def\gev{\,{\ifmmode\mathrm {GeV}\else GeV\fi}}
\def\mev{\,{\ifmmode\mathrm {MeV}\else MeV\fi}}
\def\mpl{\ifmmode \overline M_{Pl}\else $\overline M_{Pl}$\fi}
\def\cc{\ifmmode k/\overline M_{Pl}\else $k/\overline M_{Pl}$\fi}
\def\lpi{\ifmmode \Lambda_\pi\else $\Lambda_\pi$\fi}
\def\to{\rightarrow}

\def\subw{_{\rm w}}
\def\mh{\ifmmode m\sbl H \else $m\sbl H$\fi}
\def\mch{\ifmmode m_{H^\pm} \else $m_{H^\pm}$\fi}
\def\mt{\ifmmode m_t\else $m_t$\fi}
\def\mc{\ifmmode m_c\else $m_c$\fi}
\def\mz{\ifmmode M_Z\else $M_Z$\fi}
\def\mw{\ifmmode M_W\else $M_W$\fi}
\def\mws{\ifmmode M_W^2 \else $M_W^2$\fi}
\def\mhs{\ifmmode m_H^2 \else $m_H^2$\fi}
\def\mzs{\ifmmode M_Z^2 \else $M_Z^2$\fi}
\def\mts{\ifmmode m_t^2 \else $m_t^2$\fi}
\def\mcs{\ifmmode m_c^2 \else $m_c^2$\fi}
\def\mchs{\ifmmode m_{H^\pm}^2 \else $m_{H^\pm}^2$\fi}
\def\ztwo{\ifmmode Z_2\else $Z_2$\fi}
\def\zone{\ifmmode Z_1\else $Z_1$\fi}
\def\mtwo{\ifmmode M_2\else $M_2$\fi}
\def\mone{\ifmmode M_1\else $M_1$\fi}
\def\tb{\ifmmode \tan\beta \else $\tan\beta$\fi}
\def\xw{\ifmmode x\subw\else $x\subw$\fi}
\def\ch{\ifmmode H^\pm \else $H^\pm$\fi}
\def\lum{\ifmmode {\cal L}\else ${\cal L}$\fi}
\def\inpb{\,{\ifmmode {\mathrm {pb}}^{-1}\else ${\mathrm 
{pb}}^{-1}$\fi}}
\def\infb{\,{\ifmmode {\mathrm {fb}}^{-1}\else ${\mathrm 
{fb}}^{-1}$\fi}}
\def\epem{\ifmmode e^+e^-\else $e^+e^-$\fi}
\def\ppb{\ifmmode \bar pp\else $\bar pp$\fi}
\def\bsg{\ifmmode B\to X_s\gamma\else $B\to X_s\gamma$\fi}
\def\bsll{\ifmmode B\to X_s\ell^+\ell^-\else $B\to X_s\ell^+\ell^-$\fi}
\def\bstt{\ifmmode B\to X_s\tau^+\tau^-\else $B\to X_s\tau^+\tau^-$\fi}
\def\lamt{\ifmmode \tilde\lambda\else $\tilde\lambda$\fi}
\def\shat{\ifmmode \hat s\else $\hat s$\fi}
\def\that{\ifmmode \hat t\else $\hat t$\fi}
\def\uhat{\ifmmode \hat u\else $\hat u$\fi}

\newskip\zatskip \zatskip=0pt plus0pt minus0pt
\def\matth{\mathsurround=0pt}
\def\lsim{\mathrel{\mathpalette\atversim<}}

\def\atversim#1#2{\lower0.7ex\vbox{\baselineskip\zatskip\lineskip\zatskip
  \lineskiplimit 
0pt\ialign{$\matth#1\hfil##\hfil$\crcr#2\crcr\sim\crcr}}}

\def\grtsim{\,\,\rlap{\raise 3pt\hbox{$>$}}{\lower 
3pt\hbox{$\sim$}}\,\,}
\def\lsim{\,\,\rlap{\raise 3pt\hbox{$<$}}{\lower 3pt\hbox{$\sim$}}\,\,}

\def\grtsim{\,\,\rlap{\raise 3pt\hbox{$>$}}{\lower 3pt\hbox{$\sim$}}\,\,}
\def\lsim{\,\,\rlap{\raise 3pt\hbox{$<$}}{\lower 3pt\hbox{$\sim$}}\,\,}

\renewcommand{\thefootnote}{\fnsymbol{footnote}}

\begin{document} \begin{titlepage}
\rightline{\vbox{\halign{&#\hfil\cr
&SLAC-PUB-9734\cr
&May 2003\cr}}}
\begin{center}
\thispagestyle{empty}
\flushbottom

{\Large\bf Brane Localized Curvature for Warped Gravitons}
\footnote{Work supported by the Department of
Energy, Contract DE-AC03-76SF00515}
\medskip
\end{center}

\centerline{H. Davoudiasl$^{a}$, J.L. Hewett$^{b}$, and
T.G. Rizzo$^{c}$ \footnote{e-mails:
$^a$hooman@ias.edu, $^b$hewett@slac.stanford.edu, 
$^c$rizzo@slac.stanford.edu}}
\vspace{8pt} \centerline{\it $^1$School of
Natural Sciences, Institute for Advanced Study, Princeton, NJ
08540} \vspace{8pt} \centerline{\it $^2$Stanford Linear
Accelerator Center, Stanford, CA, 94309}

\vspace*{0.7cm}

\begin{abstract}

We study the effects of including brane localized curvature terms
in the Randall-Sundrum (RS) model of the hierarchy.  This leads to the 
existence of brane localized kinetic terms for the graviton. Such terms can 
be induced by brane and bulk quantum effects as well as Higgs-curvature 
mixing on the brane. We derive the modified spectrum of Kaluza-Klein (KK) 
gravitons and their couplings to 4-dimensional fields in the presence of 
these terms.  We find that the masses and couplings of the KK gravitons have 
considerable  dependence on the size of the brane localized terms; 
the weak-scale phenomenology of the model is consequently modified . 
In particular, the weak-scale spin-2 graviton resonances which generically 
appear in the RS model may be significantly  lighter than previously assumed.
However, they may avoid detection as their widths may be too narrow to be 
observable at colliders.  In the contact interaction limit, for a certain range
of parameters, the experimental reach for the scale of the theory is 
independent of the size of the boundary terms. 

\end{abstract}

\renewcommand{\thefootnote}{\arabic{footnote}} \end{titlepage}

\section{Introduction}

The Randall-Sundrum (RS) model\cite{RS1} provides a natural
geometric picture for the hierarchy between the weak scale and the
apparent gravitational scale, $\mpl$.  In this model, two 3-branes
truncate a 5-dimensional curved spacetime.  In the original RS 
construction, the Standard Model (SM) fields
reside on the SM (or TeV) brane where
a scale, which is exponentially smaller than the fundamental 5-d scale, is
induced by the geometry.  Gravity is localized on the second brane, known as 
the Planck brane. This exponential hierarchy is controlled
by the distance $r_c$ between the two branes and can generate the
weak scale on the SM brane.  The physics of the graviton
Kaluza-Klein (KK) modes is governed by the same scale. It has been
shown that $r_c$ can be stabilized by bulk
scalar fields\cite{GW}, resulting in the appearance of a single 
weak scale radion field.

The most generic and
distinctive signature of this model\cite{RS1,DHR} is the presence of TeV-scale 
spin-2 graviton KK resonances at colliders.  It is 
important to examine whether modifications arising in
a more realistic RS scenario could affect the KK graviton
phenomenology. One such modification is the introduction of brane 
curvature terms for the graviton.  These terms
respect all 4-d symmetries, and are expected to be present in a
4-d effective theory, leading to Brane Localized Kinetic Terms (BLKT's) 
for the graviton.  It has been argued that brane quantum
effects can generate BLKT's\cite{DGP,other}, and that these terms
are required as brane counter terms for bulk quantum
effects\cite{Georgietal}.  In addition, since the Higgs field is
assumed to reside on the SM brane, a Higgs-curvature mixing term
can be included in the brane action.  Such a term is allowed by
all 4-d symmetries and leads to Higgs-radion mixing with important
phenomenological consequences\cite{radionpapers,omm}.  We observe that
the same Higgs-curvature mixing term can also induce a BLKT for the
graviton on the SM brane.  Thus, there are many good theoretical
reasons for assuming the presence of such terms for gravitons.

The effects of BLKT's on graviton physics have been studied for
flat spacetimes\cite{DGP,ADDfromFermilab} and result in novel
features.  Boundary terms have also been shown to significantly modify the KK
phenomenology of bulk gauge fields in
flat\cite{Fermilab} and warped geometries\cite{DHR2,Fermilab2}.  For
the case of an alternative  RS model with only one brane and an
uncompactified 5$^{th}$ dimension, the gravitational and cosmological
effects of these terms have been analyzed\cite{Kiritsisetal}.
However, the case of the compactified RS model, which is 
relevant for weak scale
physics, in principle involves graviton boundary terms on both branes and
has yet to be examined.  We study this case here and investigate the effects 
on the phenomenology of the
weak scale KK gravitons in the presence of localized curvature terms.

We derive the modified graviton KK spectrum and
couplings and show that a significant dependence on the size of such
terms emerges.  We illustrate that this leads to considerable
changes in the low energy 4-d phenomenology of the RS model. These
changes could have important implications for future searches and
the potential discovery of model signatures.  In addition, we 
derive bounds
on the magnitude of the localized curvature terms on the Planck and
TeV branes, by requiring the absence of graviton and radion ghost
states.

The outline of this paper is as follows. 
In the next section, we derive the necessary formalism.  Our numerical 
analysis is presented in section 3 and section 4 contains some
concluding remarks.

\section{Formalism}

Here, we present our formalism and notation.  We work in the
RS background, as given by the following metric
\begin{equation}
ds^2 = e^{- 2 \sigma} \eta_{\mu \nu} d x^\mu d x^\nu +
r_c^2 d \phi^2,
\label{metric}
\end{equation}
where $\sigma = k r_c |\phi|$, $k$ is the curvature scale of order 
$\mpl$, $r_c$ is
the radius of compactification, and $\phi \in [-\pi, \pi]$. The extra 
dimension is here compactified on an $S^1/Z_2$ orbifold. The 
gravitational action $S_G$, augmented by brane localized curvature terms 
 on the Planck and TeV
branes, is assumed to be given by

\begin{equation}
S_G = \frac{M_5^3}{4}\int d^4 x \int r_c d\phi \, \sqrt {-G} \,
\left\{R^{(5)} + [g_0 \, \delta(\phi) +
g_\pi \, \delta(\phi - \pi)] \, R^{(4)} + \ldots \right\},
\label{SG}
\end{equation}
where $M_5$ is the 5-d reduced Planck mass, $G \equiv \det
(G_{\mu \nu})$ and $g_{0, \pi}$ are numerical coefficients which are 
treated as phenomenological inputs in our calculations; one
may expect $g_{0, \pi} \sim 1$ based on naturalness aguments.
Here, $R^{(5)}$ denotes the 5-d gravitational curvature and $R^{(4)}$
is the 4-d curvature which is constructed  only out of 4-d Minkowski metric
components and derivatives.  The terms proportional to
$g_{0, \pi}$ represent the only modifications to the original RS
proposal \cite{RS1} that we study in this work.
Note that we have assumed that the form of the brane terms are exactly 
the same as the bulk term except that they are calculated with the 
relevant 4-d metrics. In principle, one can imagine brane terms which are 
quadratic or even higher order in $R$, but these terms are small and 
would violate the philosophy 
of the original model where classical general relativity is applicable. 
As we will see below, brane terms of the type we consider here 
only alter the detailed 
structure while retaining the general nature of the 
usual RS solutions. 

We will consider metric perturbations of the form
\begin{equation}
G_{\mu \nu} = e^{- 2 \sigma}(\eta_{\mu \nu} +
\kappa_5 h_{\mu \nu}),
\label{pert}
\end{equation}
with $\kappa_5 = 2 M_5^{-3/2}$; $h_{\mu \nu}$ is the
5-d graviton field.  The transverse-traceless gauge
will be implemented in our calculations; this corresponds to taking  
$\partial^\mu h_{\mu \nu} = h^\mu_\mu = 0$.
We use a KK expansion for
$h_{\mu \nu}$ of the form
\begin{equation}
 h_{\mu \nu}(x, \phi) = \sum_n h_{\mu \nu}^{(n)}(x)
\frac{\chi^{(n)}(\phi)}{\sqrt{r_c}}\,,
\label{KK}
\end{equation}
where $h_{\mu \nu}^{(n)}(x)$ are the KK modes and
$\chi^{(n)}(\phi)$ are the wavefunctions along the
extra dimension.  With this expansion, Eqs.(\ref{SG})
and (\ref{pert}) imply the following equation of
motion
\begin{equation}
\frac{d}{d\phi} \left(e^{-4 \sigma} \frac{d}{d\phi}
\chi^{(n)}\right) + [1 + g_0 \, \delta(\phi) +
g_\pi \, \delta(\phi - \pi)] \,
e^{-2 \sigma} r_c^2 \, m_n^2 \, \chi^{(n)} = 0\,,
\label{EOM}
\end{equation}
where $m_n$ is the mass of the $n^{th}$ KK mode.  Away from the
boundaries at $\phi = 0, \pi$, for $n \geq 1$, we have
\cite{DHR}
\begin{equation}
\chi^{(n)}(\phi) = \frac{e^{2\sigma}}{N_n} \, \zeta_2(z_n)\,,
\label{chi}
\end{equation}
where $N_n$ are normalization constants,
$\zeta_q(z_n)\equiv J_q(z_n) + \alpha_n \, Y_q(z_n)$, with
$z_n(\phi) \equiv (m_n/k) e^\sigma$, and $\alpha_n$ are
numerical coefficients to be fixed below;
$J_q$ and $Y_q$ are Bessel functions of order $q$.

Let $\varepsilon_n \equiv z_n(0)$ and $x_n \equiv z_n(\pi)$.
Integrating Eq.(\ref{EOM}) around $\phi = 0, \pi$ yields
\begin{equation}
\zeta_1(\varepsilon_n) + \gamma_0 \, \varepsilon_n
\zeta_2(\varepsilon_n) = 0
\label{phi0}
\end{equation}
and
\begin{equation}
\zeta_1(x_n) - \gamma_\pi x_n \zeta_2(x_n) =0,
\label{phipi}
\end{equation}
respectively, with $\gamma_{0,\pi} \equiv g_{0,\pi} \, k r_c/2$. 
Addressing the hierarchy problem requires
$k r_c \sim 10$, hence, we expect $|\gamma_{0,\pi}| \lsim 10$.  
Eq.(\ref{phi0}) yields
\begin{equation}
\alpha_n = - \frac{
J_1(\varepsilon_n) + \gamma_0 \varepsilon_n J_2(\varepsilon_n)}
{Y_1(\varepsilon_n) +
\gamma_0 \varepsilon_n Y_2(\varepsilon_n)}\,.
\label{alphan}
\end{equation}
Here, we note that for $\gamma_0 \neq -1/2$
\footnote{Later we will show that $\gamma_0 = -1/2$
is not a physically allowed value.}, $\alpha_n \sim
\varepsilon_n^2$, where $\varepsilon_n \sim 10^{-15}$
for the lowest lying KK modes.  The roots
of Eq.(\ref{phipi}) give the masses $m_n = x_n k
e^{-k r_c \pi}$. Note that with $\alpha_n$ sufficiently small we can safely 
neglect the $Y_q$ components of the eigenvalue equation. This implies 
that the masses of the KK graviton tower are $\gamma_0$ independent 
unlike what happens in the corresponding case with gauge bosons in the bulk. 
It can be shown that for $\gamma_\pi \gg 1$,
an additional root $x^* \ll 1$ is obtained, where 
\begin{equation}
x^* \approx 2 \, (\gamma_\pi + 1/2)^{-1/2}.
\label{x*}
\end{equation}
This implies that there is a light mode in the spectrum for
$\gamma_\pi \gg 1$. We will show that this signals an instability in this 
background. 

Next, we will fix the normalization $N_n$ of the KK modes by
requiring
\begin{equation}
\int d \phi \, e^{-2 \sigma} [1 + g_0 \, \delta(\phi) +
g_\pi \, \delta(\phi - \pi)] \,
\chi^{(n)\, 2} = 1.
\label{diag}
\end{equation}
For $n = 0$, this equation yields the zero mode wavefunction
\begin{equation}
\chi^{(0)} = \left(\frac{k r_c}{1 + 2 \gamma_0}\right)^{1/2}, 
\label{chi0}
\end{equation}
which demonstrates that we must have $\gamma_0 > -1/2$ in order 
for the zero mode to be a physical state.
For $n \geq 1$, Eq.(\ref{diag}) gives
\begin{equation}
N_n^2 = \frac{e^{2 k r_c \pi}\zeta_2^2(x_n)}{k r_c} \,
(1 + \gamma_\pi^2 x_n^2 - 2 \gamma_\pi),
\label{Nn2}
\end{equation}
where terms of order $e^{- 2 k r_c \pi}$ have been ignored. In the 
limit that we can neglect the $\alpha_n$, this form implies that the 
graviton wavefunctions are $\gamma_0$ independent. Note that this  
normalization factor remains positive for all values of $\gamma_\pi$ 
so that this parameter is not constrained by these considerations. 

We can now obtain the couplings of the KK gravitons to the
energy momentum tensor $T_{\mu \nu}$ of the fields residing
on the TeV brane at $\phi = \pi$.  To do this, first we note
that the action (\ref{SG}) yields
\begin{equation}
\mpl^2 = \frac{M_5^3}{k}(1 + 2 \gamma_0),
\label{MPl}
\end{equation}
where $\mpl$ is the 4-d reduced Planck mass and terms of
order $e^{- 2 k r_c \pi}$ have been ignored.  The Lagrangian for
the interactions of $h_{\mu \nu}$ and the 4-d fields on the TeV
brane is given by
\begin{equation}
{\cal L} = \frac{1}{M_5^{3/2}} h_{\mu \nu}(x, \pi) T^{\mu \nu}(x).
\label{L1}
\end{equation}
Eqs.(\ref{chi}), (\ref{chi0}), (\ref{Nn2}), and (\ref{MPl})
then yield
\begin{equation}
{\cal L} = \frac{1}{\mpl} h^{(0)}_{\mu \nu}(x) T^{\mu \nu}(x)
+ \frac{1}{\Lambda_\pi} T^{\mu \nu}(x)
\sum_{n = 1}^\infty \lambda_nh^{(n)}_{\mu \nu}(x),
\label{L2}
\end{equation}
with $\Lambda_\pi \equiv \mpl e^{-k r_c \pi}$ as usual and
\begin{equation}
\lambda_n \equiv \left(\frac{1 + 2 \gamma_0}
{1 + \gamma_\pi^2 x_n^2 - 2 \gamma_\pi}\right)^{1/2}.
\label{lamn}
\end{equation}
As expected, the zero mode coupling is $\mpl^{-1}$ and for
$\gamma_0, \gamma_\pi \to 0$, the original RS model coupling for the non-zero 
mode KK states, 
$\Lambda_\pi^{-1}$ \cite{DHR}, is retrieved.  Choosing
$\gamma_\pi \neq 0$ renders the KK couplings
mode dependent, through $\lambda_n$.  We see that for $\gamma_\pi
x_n \gg 1$ these couplings have suppressions
$\sim (\gamma_\pi x_n)^{-1}$.

Before continuing, we consider if there are any further constraints 
which can be imposed upon the parameters $\gamma_{0,\pi}$. 
Though here we are concerned with the effect of brane terms on the masses and 
couplings of the KK gravitons, the RS metric yields another excitation, 
the radion, whose properties may also be modified by such terms. Using the 
results of Csaki \etal {\cite {omm}}, for example, 
the effects of the brane terms on the radion{\cite {luty}}  
can be determined since they are similar in form to those which induce 
Higgs-radion mixing. A short analysis shows that the brane terms lead to a 
multiplicative correction to 
the usual radion kinetic term, in the absence of the Higgs field, 
by a factor of $1-\gamma_\pi$; the corresponding $\gamma_0$-dependent 
contribution is found to be suppressed by powers of the exponential 
warp factor and we will ignore it in the following discussion. 
To avoid a radion ghost this implies that 
$\gamma_\pi \leq 1$, suggesting that negative or small positive values of 
$\gamma_\pi$ are allowed. This multiplicative factor can then be absorbed 
by a suitable field rescaling so that the radion field becomes canonically 
normalized. However, once the Higgs-radion mixing term is included, 
this bound becomes a bit more complex since mixing and brane term 
contributions can compensate for one another. 
{\it If} the brane localized curvature for gravitons arises from 
solely the same 
source as Higgs-radion mixing {\cite {omm,luty}} then we find that 
similar bounds 
on $\gamma_\pi$ can also be obtained. 
{\footnote {In the notation of Csaki \etal {\cite {omm}}, we find that the 
induced value of $\gamma_\pi$ is given by 
$\gamma_\pi=\xi (1+2\gamma_0)(v/\Lambda_\pi)^2$, where $v$ is the SM Higgs 
vev and $\xi$ is the radion-Higgs mixing parameter. Putting in some 
reasonable numbers, we see that the induced $\gamma_\pi$ is of order unity 
or less in magnitude.}} 
Of course, such brane terms can 
arise from many other sources so that a much wider range of values of 
$\gamma_\pi$ are certainly possible. Naturalness suggests that 
$\gamma_\pi \geq -10$ as discussed above and we will assume this soft 
bound in our discussion below.

\section{Analysis}

\begin{figure}[htbp]
\centerline{
\includegraphics[width=9cm,angle=90]{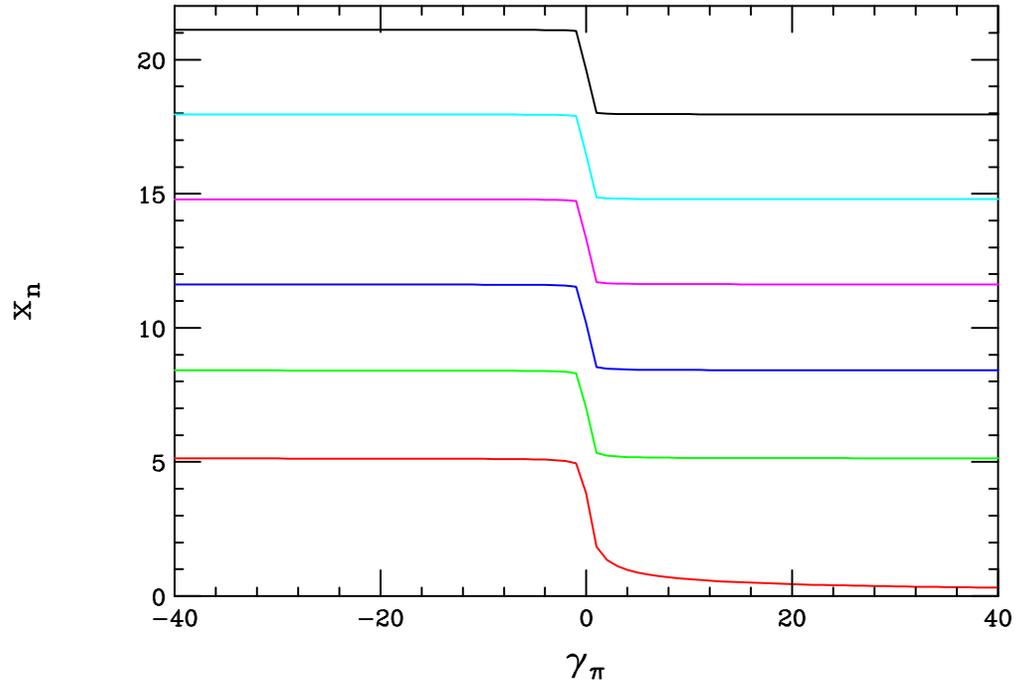}}
\vspace*{0.1cm}
\caption{The first six roots $x_{n}$ as functions of $\gamma_\pi$.  These 
results are independent of $\gamma_0$ up to terms suppressed by warp factors. 
As discussed in the text, the region $\gamma_\pi <1$ is required by the 
absence of radion ghosts taking the results of Ref.~\cite{omm}.}
\label{roots}
\end{figure}

The first step in our analysis is to examine how the mass spectrum of the 
graviton KK 
modes is influenced by the presence of the brane kinetic terms parameterized 
via $\gamma_{0,\pi}$. As was noted above, both the root equation and the 
wave functions for the graviton KK modes in the extra dimension are found to 
be 
independent of $\gamma_0$ when powers of the warp factor, $e^{-\pi kr_c}$, are 
neglected; this implies that the $x_n$ for the range of interest to us 
are essentially $\gamma_0$ independent. Recall that the masses of the KK 
states are now given by $m_n=x_n(\gamma_\pi)ke^{-kr_c\pi}$. 
Fig.~\ref{roots} displays these roots 
as a functions $\gamma_\pi$. (For purposes of comparison we note that the 
first root in the usual RS model without brane terms is given by 
$x_1 \simeq 3.83$.)  
For all KK modes, other than the case $n=1$, the $x_n$ are 
essentially constant away from the transition region near $\gamma_\pi=0$. As 
we see from 
the root equation above, once $|\gamma_\pi|$ gets large the $x_n$ are 
essentially just the roots of 
the equation $J_2(x_n)=0$ which accounts for their mostly  
$\gamma_\pi$ independent behaviour. The root 
$x_1$ behaves somewhat differently and 
scales as $\sim 2/\sqrt {\gamma_\pi}$ for large positive $\gamma_\pi$ as 
seen from Eq.(10) above. Note 
that this causes the ratio of the first two roots $x_2/x_1$ to grow  
substantially in the large and positive  
$\gamma_\pi$ region.

\begin{figure}[htbp]
\centerline{
\includegraphics[width=9cm,angle=90]{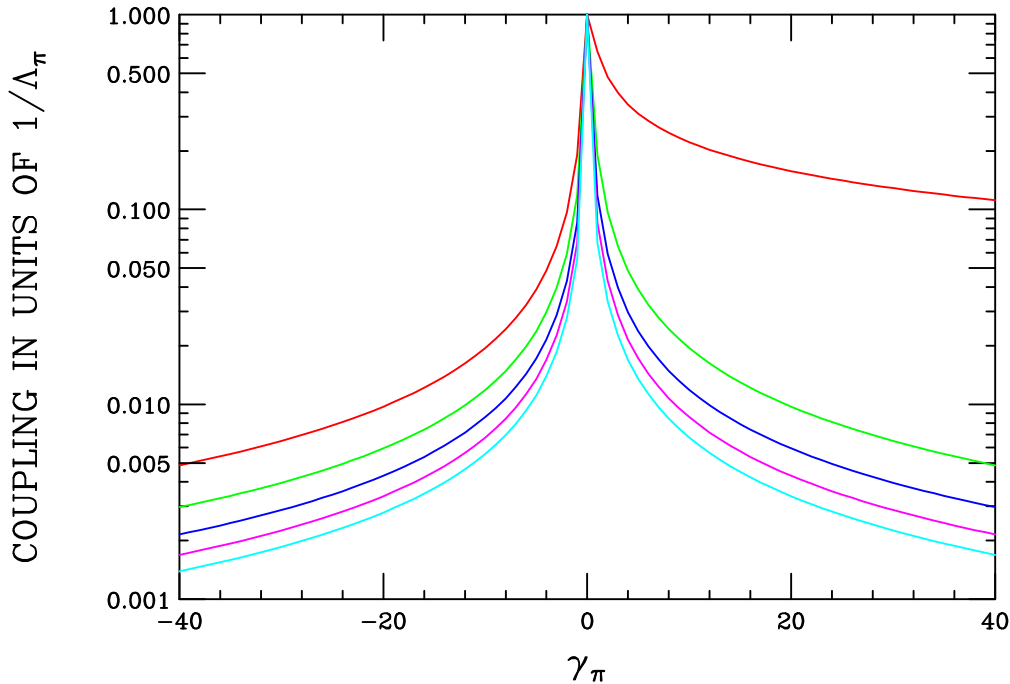}}
\vspace*{0.1cm}
\caption{The couplings of the first five KK graviton tower 
modes as functions of 
$\gamma_\pi$ assuming $\gamma_0=0$. The top curve corresponds to the 
lightest KK state with the more massive tower 
modes corresponding to the subsequently lower curves. The same restrictions 
on $\gamma_\pi$ apply as in the previous figure.}
\label{couplings}
\end{figure}

Once the $x_n$ are known the effective couplings of the KK modes to 
SM matter on the TeV brane can be immediately calculated and are shown in 
Fig. ~\ref{couplings} for the 
case $\gamma_0=0$; for other values of $\gamma_0$ 
these results are scaled by a factor of $\sqrt {1+2\gamma_0}$. 
Except for the lowest KK mode when $\gamma_\pi >0$, the 
couplings of all KK modes are observed to fall off drastically as the 
magnitude of $\gamma_\pi$ increases, demonstrating the $1/|x_n\gamma_\pi|$ 
behavior, as expected. Clearly, the 
reduced couplings of these graviton 
KK modes will have a important impact on the ability 
to detect these states either directly or indirectly at colliders. This will 
be true in particular in the case of of positive $\gamma_\pi$, where  
as we saw above the second KK mode is significantly heavier than the first  
mode and has a much 
weaker coupling to the SM brane fields. However, we note that 
the absence of a radion ghost would 
exclude this extreme situation if we apply the analysis as 
given in Ref.~\cite{omm}.

\begin{figure}[htbp]
\centerline{
\includegraphics[width=9cm,angle=90]{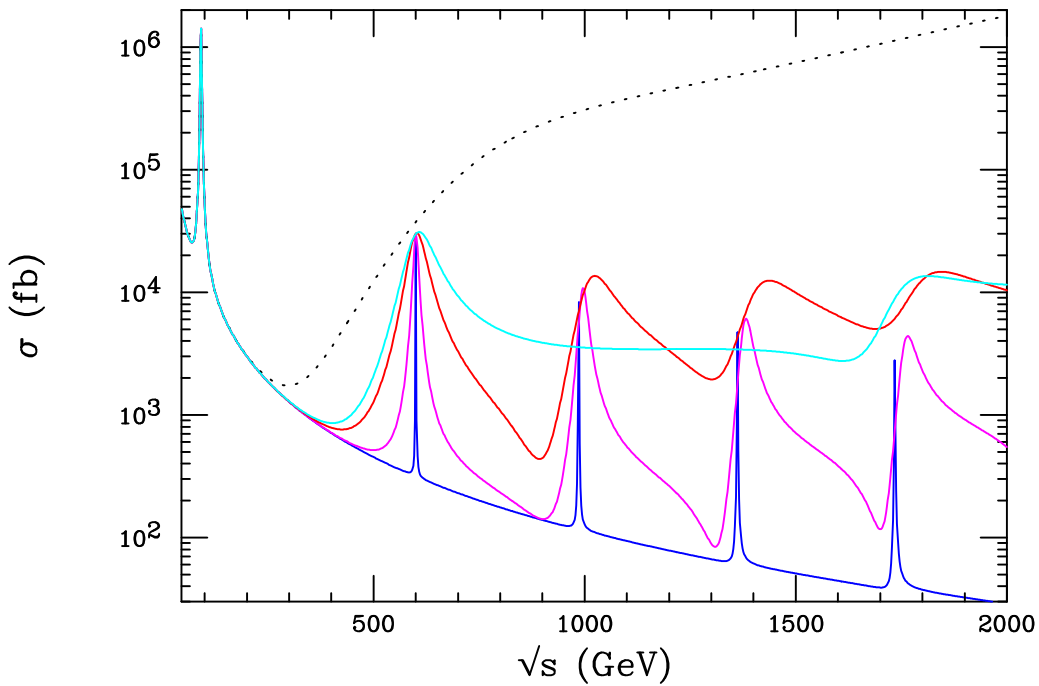}}
\vspace*{0.1cm}
\caption{Cross section for $e^+e^-\to \mu^+\mu^-$ at a linear collider 
assuming $m_1=600$ GeV and $k/\mpl=1$. The dotted curve is the standard RS 
result while from top to bottom on the right-hand side the curves correspond 
to $\gamma_\pi=1(-1,-2,-10)$ represented by cyan(red, 
magenta, blue). In all cases $\gamma_0=0$ 
is assumed.}
\label{bumps1}
\end{figure}

\begin{figure}[htbp]
\centerline{
\includegraphics[width=9cm,angle=90]{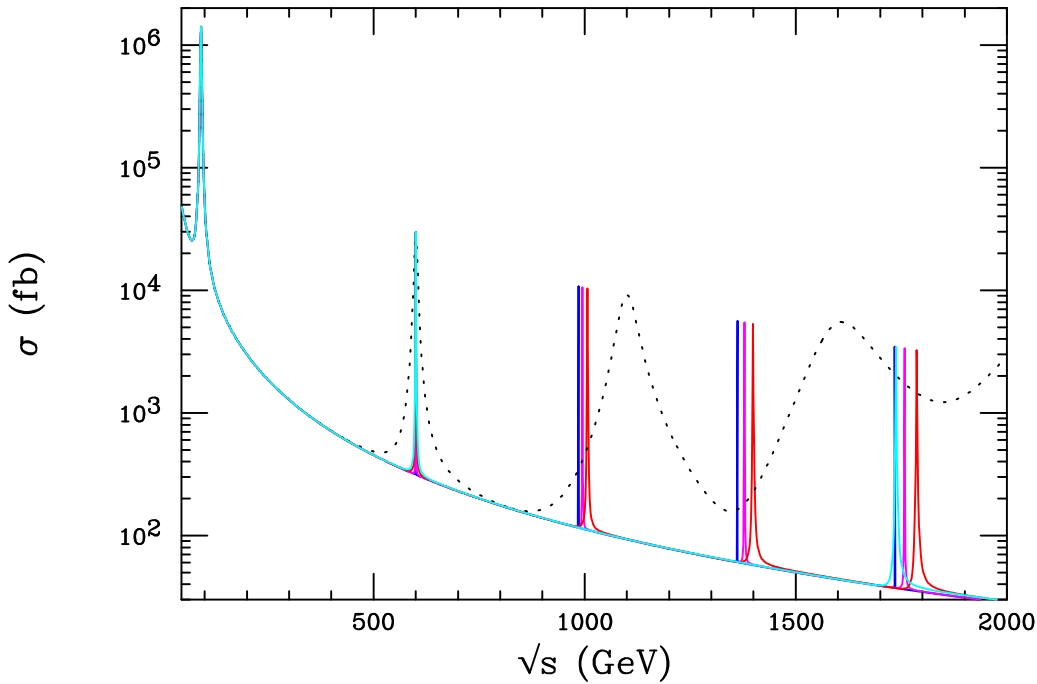}}
\vspace*{0.1cm}
\caption{Same as the previous figure but now for $k/\mpl=0.1$. The 
rapidly rising curve is the usual RS scenario prediction. The resonances 
for the case $\gamma_\pi=-10$ have become essentially a spike here due to the 
resolution in our plotting routine.}
\label{bumps2}
\end{figure}

Perhaps the most important experimental signature for the RS model is the 
appearance of spin-2 KK resonances at colliders. As shown above, 
when $\gamma_\pi$ is non-zero 
the masses and couplings of these KK states are altered from the usual 
expectations for which the phenomenology has by now been well explored. 
Figs.~\ref{bumps1} and ~\ref{bumps2} 
show some of the possible modifications in the presence of the 
boundary terms to the production cross section 
and KK spectrum for the cases $k/\mpl=1$ and 0.1, 
respectively, for the process $e^+e^- \to \mu^+\mu^-$ at a linear collider 
assuming the mass of the first KK resonance is 600 GeV. For the case of 
$k/\mpl=1$ in the usual RS model, which corresponds to $\gamma_\pi=0$ and is 
represented by the dotted curve in 
Fig.~\ref{bumps1},  the KK states are sufficiently strongly coupled to smear 
out the individual resonance structures and produce only a large shoulder in 
the cross section. For non-zero $\gamma_\pi$, there is a significant change 
in these expectations due to the reduced values of the 
couplings and the additional mass 
spectrum modifications. This is seen explicitly in these two 
figures. Note that the widths of these graviton resonances scale as 
$\Gamma_n/m_n \sim (x_n \lambda_n k/\mpl)^2$ where $\lambda_n$ is the 
relative coupling 
strength for the $n$-th KK mode calculated in Eq.(17) and 
shown in Fig.~\ref{couplings}. For positive 
values of $\gamma_\pi$ the lowest KK mode is well separated 
in mass from the next more massive state. 
For $\gamma_\pi <0$ the strong constructive interference between 
the resonances rapidly vanishes and individual peaks become
observable.  As $\gamma_\pi$  decreases further this results in a series 
of very narrow, spike-like  peaks once values of $\sim -10$ are reached.

When $k/\mpl=0.1$ the effects of the brane terms are in some sense much more 
severe as can be seen from Fig.~\ref{bumps2} due to the 
further reduction in the 
resonance widths by a factor of 100. The usual resonances, the first 
few of which are reasonably narrow in the standard RS scheme in this case, 
are converted into 
narrow spikes some if not all of which will yield event rates which are 
quite small compared to the continuum background in any given mass bin. 
For $\gamma_\pi$ positive and of order unity the unique feature here is 
the large mass gap between the first and second resonances. 
For $\gamma_\pi$ negative the resonance structures are also narrow but with 
a smaller mass gap. 
With this or even smaller values of $k/\mpl$, it may be difficult to observe 
anything, even the first KK graviton excitation, over much of the range of 
$\gamma_\pi$ unless use is made of radiative returns or one is lucky 
choosing the collider center of mass energy in performing a scan. It is 
clear from our discussion that it is important to see at least two of these 
resonance peaks to uniquely identify the model and extract out the relevant 
parameters. 

\begin{figure}[htbp]
\centerline{
\includegraphics[width=9cm,angle=90]{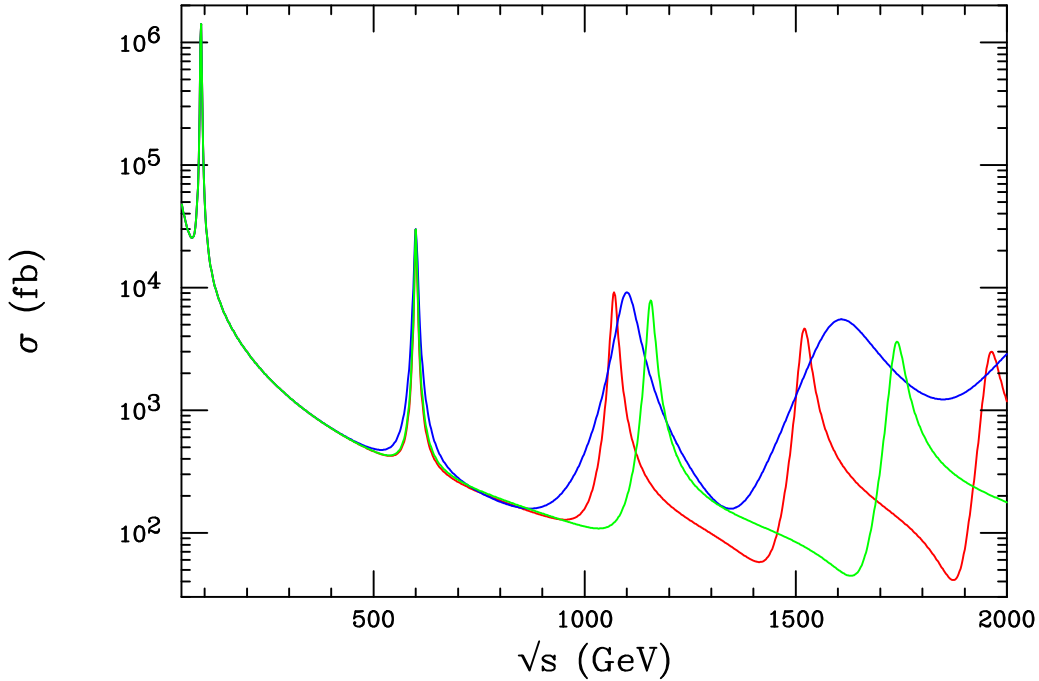}}
\vspace*{0.1cm}
\caption{Same as the previous figure with $k/\mpl=0.1$ but now for 
$\gamma_\pi=-0.2, 0, $ and 0.2 corresponding to the red, blue and green 
curves, respectively, which are represented by the middle, top and bottom 
curves on the right-hand side.}
\label{new2}
\end{figure}

The shifts in the mass spectrum and couplings of the KK resonances induced 
by the brane kinetic terms are so large that even for values of 
$|\gamma_\pi|=0.2$, there are substantial modifications to the shape of 
the production 
cross section. This is seen explicitly in Fig.~\ref{new2}.

\begin{figure}[htbp]
\centerline{
\includegraphics[width=9cm,angle=90]{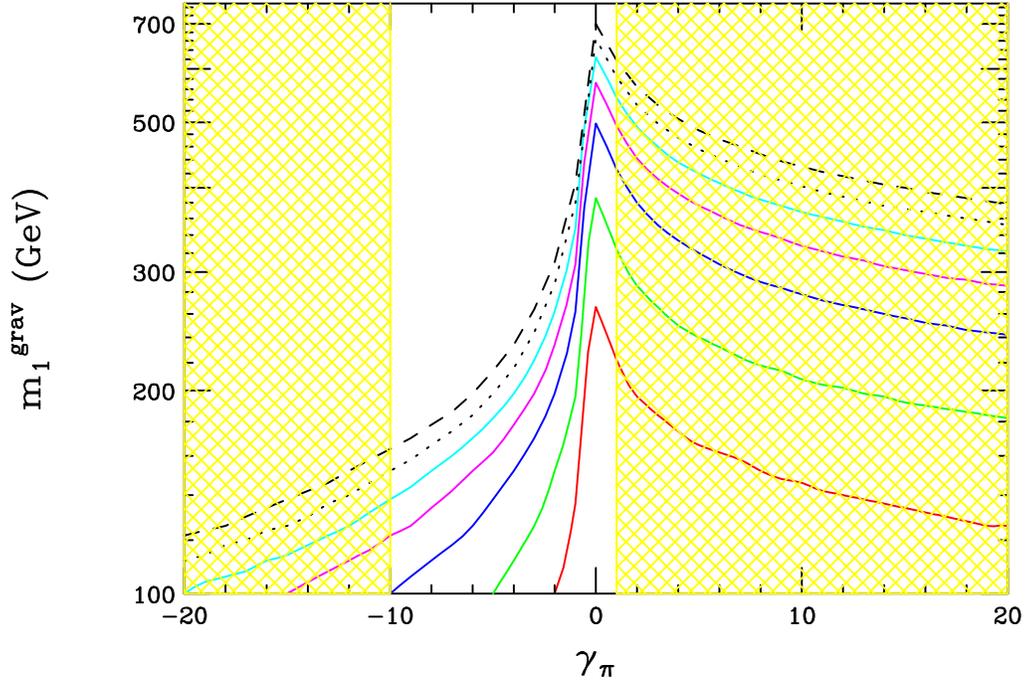}}
\vspace*{0.1cm}
\caption{Search reach for the first graviton KK resonance employing the 
Drell-Yan channel at Tevatron Run I as 
a function of $\gamma_\pi$ assuming $\gamma_0=0$. From bottom to 
top on the RHS 
of the plot, the curves correspond to $k/\mpl=0.01, 0.025, 0.05, 0.075, 0.10, 
0.125$ and 0.15, respectively. The unshaded region is that allowed by 
naturalness considerations and the requirement of a ghost-free radion 
sector.}
\label{search1}
\end{figure}

\begin{figure}[htbp]
\centerline{
\includegraphics[width=9cm,angle=90]{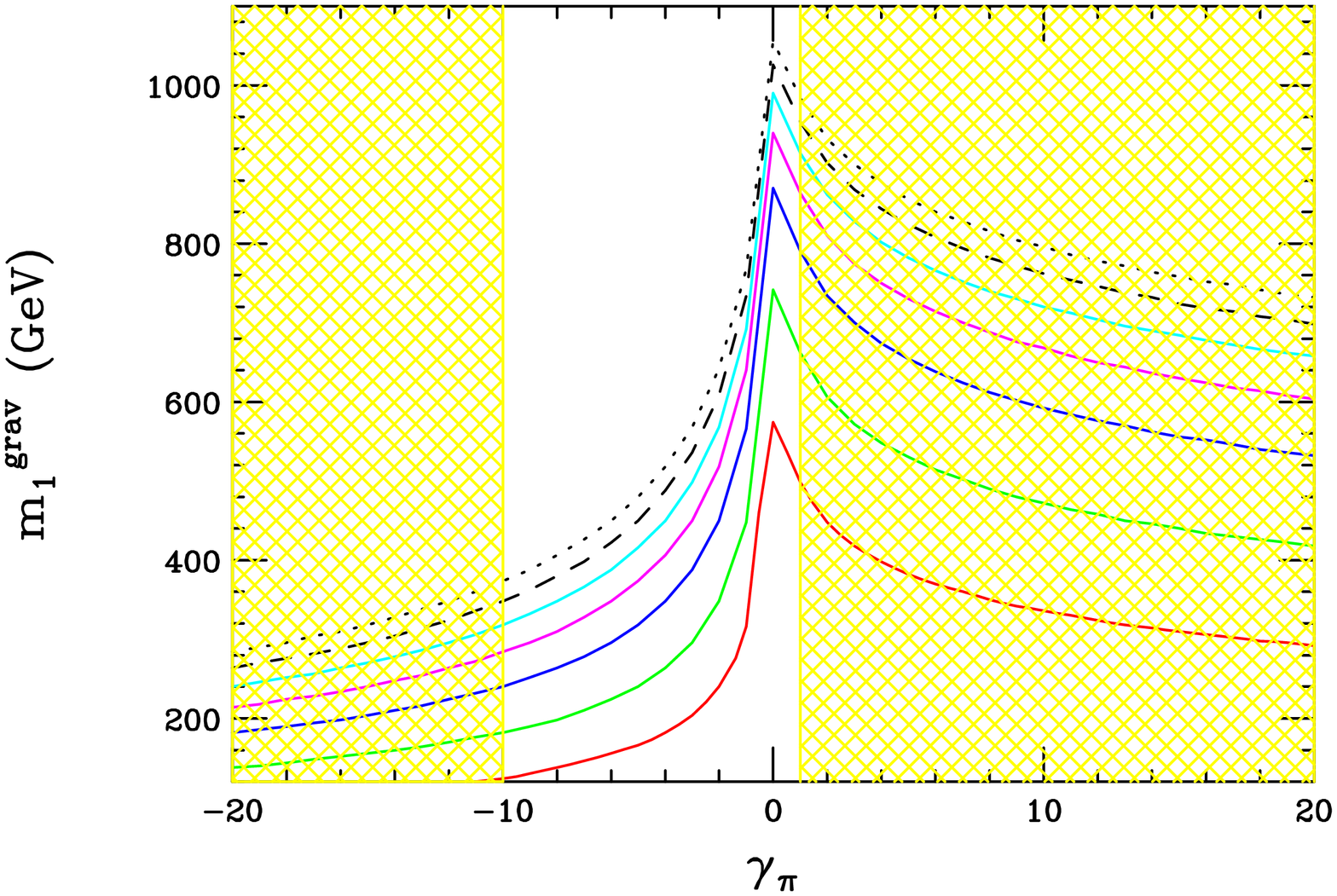}}
\vspace*{0.1cm}
\caption{Same as the previous figure but now for Run II, $\sqrt s=1.96$ TeV,  
Tevatron assuming an integrated luminosity of 5 $fb^{-1}$.}
\label{search2}
\end{figure}

\begin{figure}[htbp]
\centerline{
\includegraphics[width=9cm,angle=90]{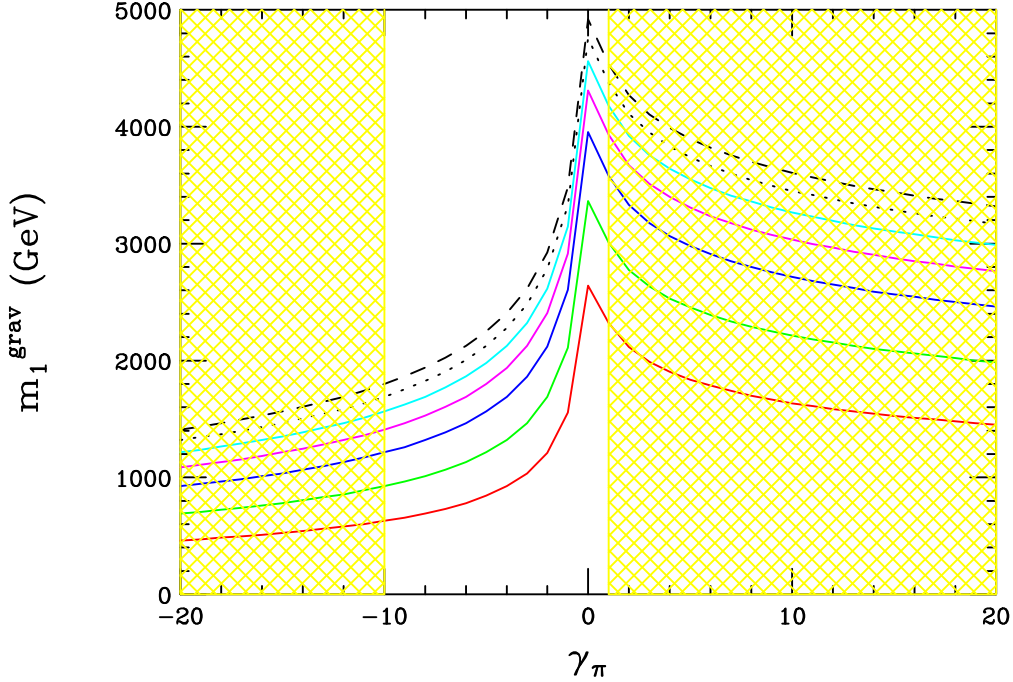}}
\vspace*{0.1cm}
\caption{Same as the previous figure but now for the LHC with an integrated 
luminosity of 100 $fb^{-1}$.}
\label{search3}
\end{figure}

If the graviton KK modes are kinematically accessible they will most likely 
be first observed at a hadron collider. As discussed in our earlier work, 
the lack of observation of KK graviton resonances in, \eg, the Drell-Yan 
channel at the Tevatron already places a constraint on the RS model parameter 
space{\cite {DHR}}. This analysis can be modified in order to examine the 
influence of brane kinetic terms for the graviton; the 
results are presented in 
Figs.~\ref {search1}, ~\ref{search2} and ~\ref{search3} keeping in mind that 
values of $k/\mpl \lsim 0.1$ are theoretically 
preferred. These figures show the 
search reach for the first graviton KK resonance in Drell-Yan production 
as a function 
of $\gamma_\pi$ for different values of $k/\mpl$.

The search reach from Run I of the Tevatron, which constitutes the present 
lower bound on the mass of the first KK graviton excitation, 
is shown in Fig.~\ref{search1}. Here we see that, as expected, the search 
reach degrades quite rapidly as we move 
away from the canonical value of $\gamma_\pi=0$. 
Note that for negative values of $\gamma_\pi$ it is still 
quite possible for the mass of this lightest state to be less than 200 GeV 
(provided this is not excluded by LEP searches for narrow resonances). 
For negative values of $\gamma_\pi$ the corresponding bound on $\Lambda_\pi$ 
can be approximately obtained from the $m_1$ constraint by dividing by 
$\simeq 5k/\mpl$.  

Fig.~\ref{search2} shows that future running of the Tevatron 
during Run II can be useful in covering some of the 
large $|\gamma_\pi|$, small KK mass region. The higher energy and integrated 
luminosity will result in a reasonable increase in the overall parameter space 
coverage compared to Run I but, as can be seen from the figure, 
will still allow for the existence of very light KK gravitons.

Fig.~\ref{search3} shows the potential search reach for the first KK 
graviton resonance at the LHC. Even in this case 
one sees that there is a significant degradation in the reach away from 
$\gamma_\pi=0$. Unlike in the standard RS scenario 
without brane terms we see that 
the LHC can no longer cover all of the interesting parameter space for this 
model due to the reduced coupling of the first KK resonance. For example we 
see from the figure that for large negative $\gamma_\pi$, a first KK 
graviton resonance with a mass of 600 GeV and $k/\mpl=0.01$ may still 
miss detection - even at the LHC.  

\begin{figure}[htbp]
\centerline{
\includegraphics[width=9cm,angle=90]{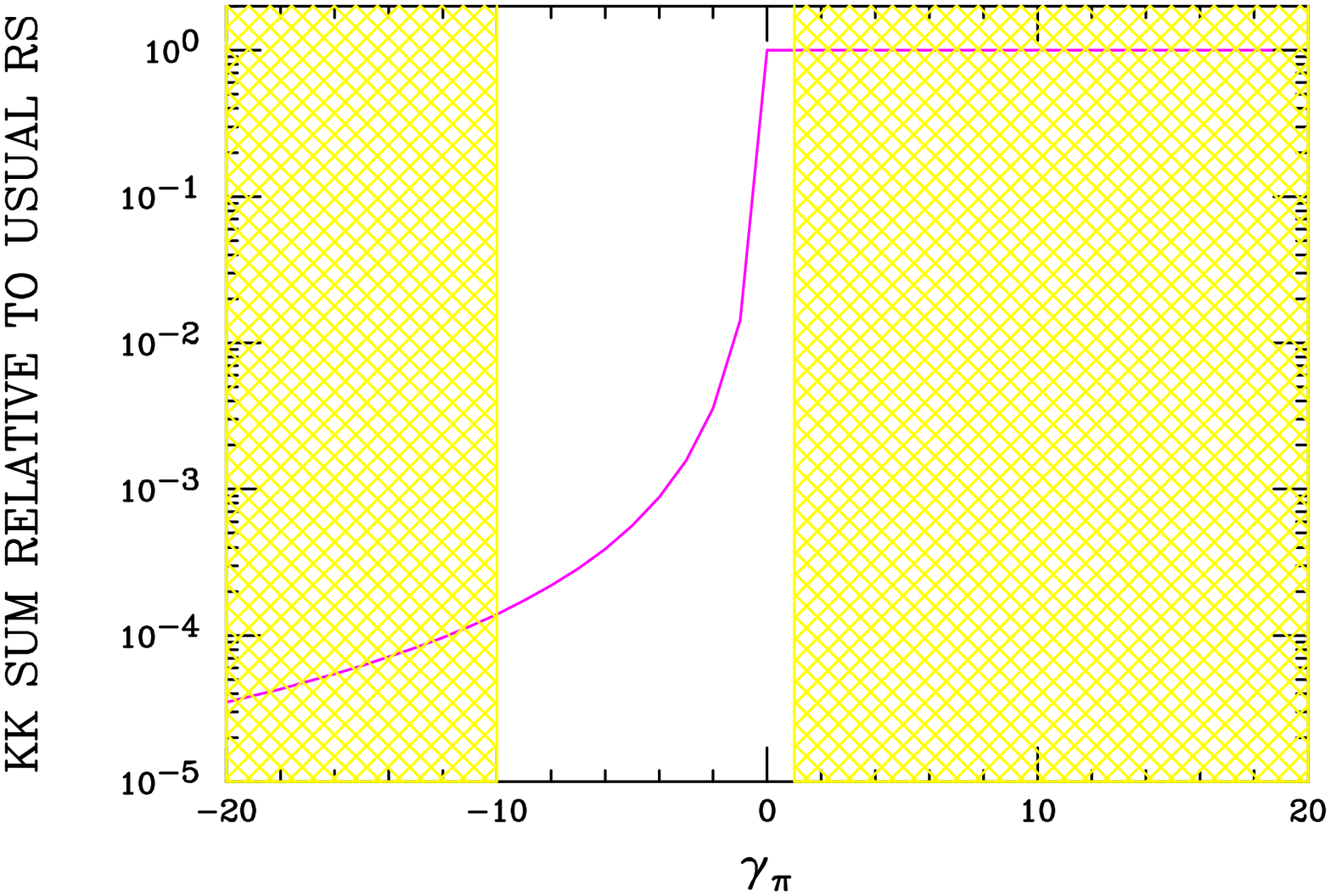}}
\vspace*{0.1cm}
\caption{Sum over the graviton KK towers weighted by the appropriate coupling 
strength in comparison to that obtained in the standard RS model without 
brane kinetic terms as a function of $\gamma_\pi$ assuming $\gamma_0=0$.}
\label{sum}
\end{figure}

\begin{figure}[htbp]
\centerline{
\includegraphics[width=9cm,angle=90]{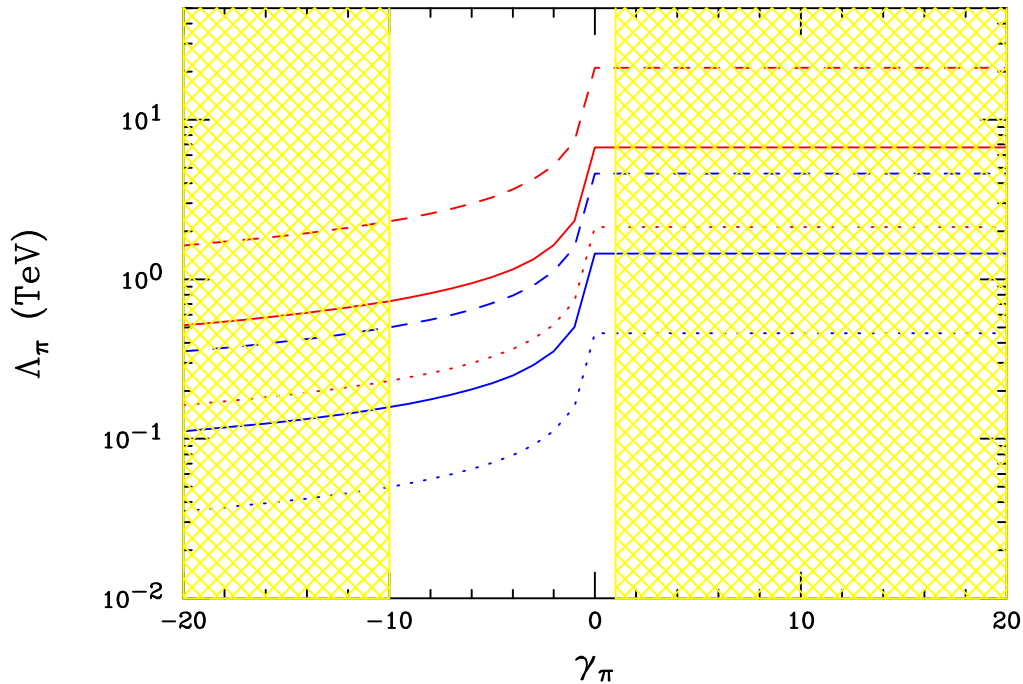}}
\vspace*{0.1cm}
\caption{Contact interaction constraints on the scale $\Lambda_\pi$ from 
the run II Tevatron with an integrated  luminosity of 2 
$fb^{-1}$(blue lower set of curves) and the LHC (red higher set of curves) 
with an integrated luminosity of 100 $fb^{-1}$. 
The dashed(solid, dotted) curves correspond to $k/\mpl=0.01(0.1, 1)$, 
respectively. The shading is as in the previous figures.}
\label{cont1}
\end{figure}

\begin{figure}[htbp]
\centerline{
\includegraphics[width=9cm,angle=90]{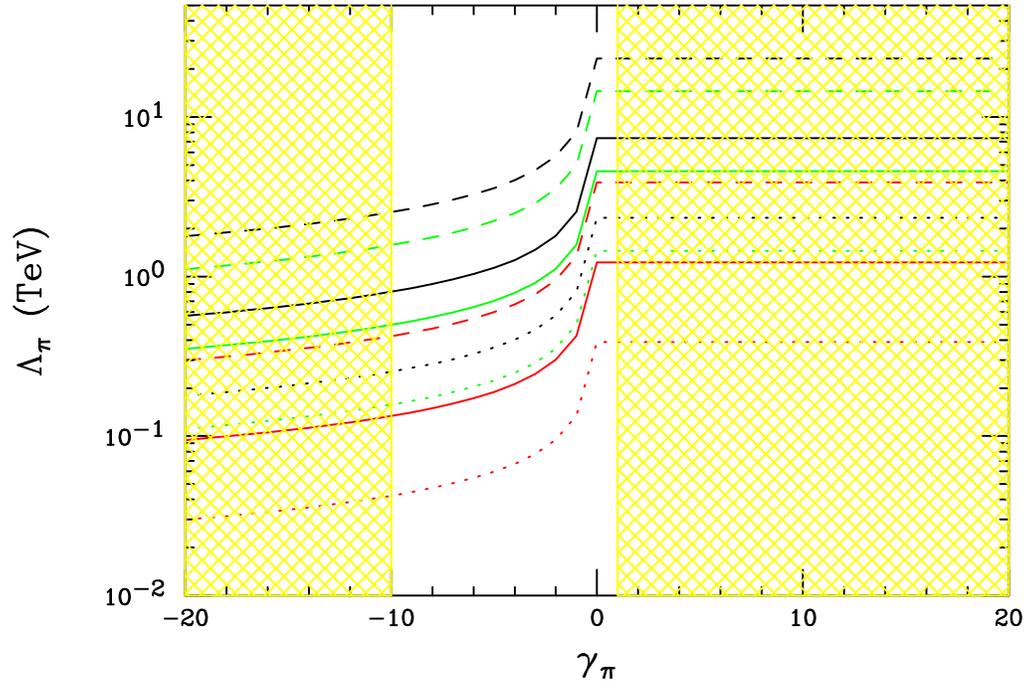}}
\vspace*{0.1cm}
\caption{Same as the previous figure but now for LEPII(red lower set of 
curves), a 500 GeV 
Linear Collider (green, middle set), or a 1 TeV Linear Collider (black, 
higher set). The LC 
integrated luminosity is assumed to be 500 $fb^{-1}$.}
\label{cont2}
\end{figure}

It is possible that the indirect effects of graviton KK tower exchange
in particle pair production may be observed.  This may be 
more important than direct graviton searches in some regions of 
parameter space.  These indirect effects are similar
to those of contact interactions, except that the graviton exchange
operator is dimension-8.  Graviton virtual effects have 
been computed for the case
of large extra dimensions in Ref. \cite{jlh}, where it was noted
that the angular distributions of the final state
particles may reveal the spin-2 nature of the indirect graviton
exchange.  Here, the 4-fermion
matrix element is obtained from that for the large extra dimensions
scenario with the replacement
\begin{equation}
{\lambda\over M^4_H} \to {\mpl^2\over 8k^2\Lambda_\pi^4}
\sum_n {\lambda_n(\gamma_\pi,\gamma_0)^2\over x_n(\gamma_\pi)^2}\,,
\end{equation}
in the limit of $m_n^2\gg s$.  In this model the sum rapidly converges.
The coupling weighted sum over the KK tower exchanges differs
from the conventional RS model by the ratio
\begin{equation}
{\cal R}= 
{\sum_n {\lambda_n(\gamma_\pi,\gamma_0)^2\over {x_n(\gamma_\pi)
^2}}\over {\sum_n {\lambda_n(0)^2\over {x_n(0)^2}}}}\,,
\end{equation}
which is displayed in Fig. \ref{sum}.  We see from the figure
that for positive values of $\gamma_\pi$ the effects of the
suppressed couplings and the modified roots exactly compensate
each other.  The standard RS predictions are thus maintained
in this region.  The contact interaction bounds on $\Lambda_\pi$
obtained from the lepton pair invariant mass spectrum and 
forward-backward asymmetry in Drell-Yan production at
the Tevatron and LHC  are presented in Fig. \ref{cont1} as a 
function of $\gamma_\pi$.  The results from unpolarized
(and polarized in the case of the Linear Collider) angular
distributions, summing over $e\,,\mu\,,\tau\,,c\,,b$ (and $t$ if
kinematically accessible) final states, as well as $\tau$
polarization distributions are shown in Fig. \ref{cont2}  for LEP II
and the Linear Collider.  From these figures we see that the
constraints from 4-fermion interactions are rather weak for 
negative values of $\gamma_\pi$, even for the higher energy
colliders.   The searches from both direct graviton production at
hadron colliders and the indirect exchange of graviton KK states
in fermion pair production are thus difficult
in this parameter space region and low mass graviton states
may escape detection.  It is possible that searches for narrow
s-channel resonances in $e^+e^-$ collisions or direct
graviton production in $e^+e^-\to\gamma G$ may extend
the discovery reach in this region.


\section{Conclusions}

The most generic and distinctive signature of the RS model of the hierarchy
is the appearance of weak scale, spin-2 KK resonances of the 5-d
graviton.  The discovery of such resonances at
a collider will provide strong evidence for the RS picture.  It is
therefore important to understand how the spectrum and the
couplings of these states could be affected by well-motivated
theoretical modifications of the original proposal.  The addition
of brane localized curvature terms is such a modification.  These
terms yield contributions to the graviton BLKT's and can be
generated from brane and bulk quantum effects.  They can also arise
at tree level due to Higgs-curvature mixing.

In this paper, we studied the consequences of adding brane
curvature terms in the RS model.  We derived the modified spectrum
and couplings of the graviton KK states to the fields on the SM brane.
The presence of localized curvature on the SM brane leads to
important changes in the weak scale phenomenology.  
Requiring that the effective theory is free of KK and radion
ghosts places lower and upper bounds on the coefficients
$\gamma_0$ and $\gamma_\pi$ of these terms on the Planck and SM
branes, respectively.

We showed that, within the natural and allowed values of these
coefficients, the variation in the shape and production cross
sections of the resonances is significant. In particular, the KK 
resonances become very narrow for some regions of the parameter 
space and may be too narrow to be observed at colliders. We found that  
very light KK gravitons, of order 200-600 GeV may escape 
detection at the Tevatron and LHC. In the contact
interaction limit, where the KK masses are too large for direct
production, we observed that the collider reach for the SM scale
$\Lambda_\pi \sim$ TeV varies considerably with $\gamma_\pi < 0$.
However, for $\gamma_\pi > 0$, this reach remains constant. 
The cause of this behavior
is the interplay of the modified masses and couplings of
the KK states. In brief, we
found that the modifications resulting from brane localized
curvature terms in the RS model have interesting features and that
they are important for future phenomenological studies and
experimental searches.

\noindent{\bf Acknowledgements}

One of us (H.D.) would like to thank G. Kribs and C. Wagner 
for discussions related to this work. The work of
H.D. was supported by the US Department of Energy under contract
DE-FG02-90ER40542.

%
\def\MPL #1 #2 #3 {Mod. Phys. Lett. {\bf#1},\ #2 (#3)}
\def\NPB #1 #2 #3 {Nucl. Phys. {\bf#1},\ #2 (#3)}
\def\PLB #1 #2 #3 {Phys. Lett. {\bf#1},\ #2 (#3)}
\def\PR #1 #2 #3 {Phys. Rep. {\bf#1},\ #2 (#3)}
\def\PRD #1 #2 #3 {Phys. Rev. {\bf#1},\ #2 (#3)}
\def\PRL #1 #2 #3 {Phys. Rev. Lett. {\bf#1},\ #2 (#3)}
\def\RMP #1 #2 #3 {Rev. Mod. Phys. {\bf#1},\ #2 (#3)}
\def\NIM #1 #2 #3 {Nuc. Inst. Meth. {\bf#1},\ #2 (#3)}
\def\ZPC #1 #2 #3 {Z. Phys. {\bf#1},\ #2 (#3)}
\def\EJPC #1 #2 #3 {E. Phys. J. {\bf#1},\ #2 (#3)}
\def\IJMP #1 #2 #3 {Int. J. Mod. Phys. {\bf#1},\ #2 (#3)}
\def\JHEP #1 #2 #3 {J. High En. Phys. {\bf#1},\ #2 (#3)}

\end{document}